\documentclass[conference]{IEEEtran}
\usepackage{cite}
\usepackage{amsmath,amssymb,amsfonts}
\usepackage{algorithmic}
\usepackage{graphicx}
\usepackage{textcomp}
\usepackage{xcolor}
\usepackage[short]{optidef}
\usepackage{tikz}
\usetikzlibrary{arrows,arrows.meta,shapes,positioning,shadows,trees}
\usepackage{siunitx}
\usepackage{lipsum}
\usepackage{systeme}

\makeatletter
\newcommand*\titleheader[1]{\gdef\@titleheader{#1}}
\AtBeginDocument{%
  \let\st@red@title\@title
  \def\@title{%
    \bgroup\normalfont\large\centering\@titleheader\par\egroup
    \vskip1.5em\st@red@title}
}
\makeatother

\newcommand{\includePDF}[2]
{
\IfFileExists{Figures_Cropped/#2.pdf}{}{\immediate\write18{pdfcrop Figures_Uncropped/#2.pdf Figures_Cropped/#2.pdf}}
\includegraphics[#1]{Figures_Cropped/#2.pdf}
}

\NewDocumentCommand{\vartwo}{ >{\SplitArgument{2}{,}}m }{ \finalvartwo#1 }
\NewDocumentCommand{\finalvartwo}{mmm}{ \ensuremath{#1_\text{#2}} }

\NewDocumentCommand{\varthree}{ >{\SplitArgument{3}{,}}m }{ \finalvarthree#1 }
\NewDocumentCommand{\finalvarthree}{mmmm}{ \ensuremath{#1_{\text{#2},#3}} }

\def\BibTeX{{\rm B\kern-.05em{\sc i\kern-.025em b}\kern-.08em
    T\kern-.1667em\lower.7ex\hbox{E}\kern-.125emX}}

\begin{document}

\onecolumn

© 2021 IEEE.  Personal use of this material is permitted.  Permission from IEEE must be obtained for all other uses, in any current or future media, including reprinting/republishing this material for advertising or promotional purposes, creating new collective works, for resale or redistribution to servers or lists, or reuse of any copyrighted component of this work in other works.

\newpage

\twocolumn
\title{\color{blue}Convex Optimization of Speed and Energy Management System for Fuel Cell Hybrid Trains}

\author{\IEEEauthorblockN{Rabee Jibrin, Stuart Hillmansen, Clive Roberts, Ning Zhao}
\IEEEauthorblockA{\textit{Department of Electronic, Electrical and Systems Engineering} \\
\textit{University of Birmingham}\\
Birmingham, United Kingdom \\
\{rxj956 s.hillmansen c.roberts.20 n.zhao\}@bham.ac.uk}
\and
\IEEEauthorblockN{Zhongbei Tian}
\IEEEauthorblockA{\textit{Department of Electrical Engineering and Electronics} \\
\textit{University of Liverpool}\\
Liverpool, United Kingdom \\
zhongbei.tian@liverpool.ac.uk}}


\maketitle

\begin{abstract}
We look into minimizing the hydrogen fuel consumption of hydrogen hybrid trains by optimizing their operation. The powertrain considered is a fuel cell charge-sustaining hybrid. Convex optimization is utilized to compute optimal speed and energy management trajectories. The barrier method is used to solve the optimization problems quickly on the order of tens of seconds for the entire journey. Simulations show a considerable reduction in fuel consumption when both trajectories---speed and energy management---are optimized concurrently within a single optimization problem in comparison to being optimized separately in a sequential manner---optimizing energy management after optimizing speed. It is concluded that the concurrent method greatly benefits from its holistic powertrain knowledge while optimizing all trajectories together within a single optimization problem. 
\end{abstract}


\section{Introduction}

\subsection{Motivation}

Hydrogen trains are expected to play a role in decarbonizing the railways \cite{RN243}. While hydrogen trains are virtually emissions free at point-of-use and can achieve a driving range similar to that of diesel trains, their total cost of ownership is currently higher primarily due to the cost of hydrogen fuel \cite{RN200}; therefore, it becomes imperative to reduce hydrogen consumption in order to improve competitiveness.

We aim at reducing the fuel consumption of hydrogen hybrid trains by optimizing their speed and energy management system. We achieve this by formulating an optimization problem tailored to the intricacies of the powertrain. The focus herein is on the polymer electrolyte membrane fuel cell (PEMFC) in a charge-sustaining series hybrid configuration. The PEMFC is widely regarded for its high technology readiness level for transport applications \cite{RN822}.

\subsection{Background}

The 2019 IEEE VTS Motor Vehicles Challenge brought attention to the energy management system (EMS) of fuel cell hybrid locomotives \cite{RN876}. The EMS determines power distribution among multiple power-sources and is thus a vital determinant of hybrid vehicle efficiency. Several approaches have been used for fuel cell hybrid EMS, such as: fuzzy logic \cite{RN507}, the equivalent consumption minimization strategy \cite{RN687}, and model predictive control \cite{RN571}. An extensive body of EMS literature for fuel cell hybrid vehicles has been reviewed by \cite{RN318}. Simulations suggest that optimization-based algorithms outperform their rule-based counterparts \cite{RN565}. 

The aforementioned only deals with the EMS given a reference speed profile to follow, which is usually generated \textit{a priori} by a separate speed optimization algorithm, e.g., \cite{RN623} optimize the speed of non-hybrid hydrogen trains. This sequential approach to optimizing speed and EMS separately (also known as multi-layer) is the most common in literature owing to its simplicity but is potentially sub-optimal, since the former speed optimization step lacks control over the EMS trajectory as well as complete knowledge of powertrain characteristics solely known by the latter EMS optimization step. Concurrent formulations that incorporate all powertrain knowledge and optimize both speed and EMS within the same optimization problem bypass this impediment but often rely on computationally expensive techniques that cannot be deployed in real-time, e.g., dynamic programming \cite{RN912,RN616} and indirect optimal control \cite{RN613,RN909}. Relaxed convex formulations that are computationally lighter have been proposed but attain the true optimal solution under certain conditions \cite{RN776,RN886}.

Concurrent optimization of speed and EMS was also considered for catenary-powered trains supported by on-board energy storage \cite{RN924}. It was shown therein that knowledge of energy storage power constraints at time of speed optimization is crucial for optimality and feasibility. The same group of authors expanded their work from optimizing between a single pair of stations to a typical rail journey with multiple stops \cite{RN925}, but integer programming was used and the solution had state-of-charge discontinuities at station stops.

\subsection{Contribution}

Current literature lacks a computationally light method to concurrently optimize the speed and EMS of hybrid trains. We aim to alleviate this by formulating a relaxed convex optimization problem that remains optimal under less restrictive conditions than previous relaxed formulations. Furthermore, the battery's state-of-charge is optimized for the entire journey without any discontinuities at station stops.

\subsection{Outline}

After introducing necessary mathematical models in section II, the concurrent formulation and a benchmark sequential formulation are presented in section III. The methods are compared using simulation results in section IV. Conclusions and avenues for further work are mentioned in section V.

\section{Modeling}

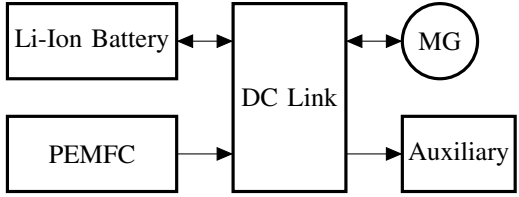
\begin{figure}[!t]
\centering
  \begin{tikzpicture}
  \draw[black, very thick] (-0.25,0) rectangle (2,1) node[pos=0.5] {Li-Ion Battery};
  \draw[black, very thick] (-0.25,-1.5) rectangle (2,-0.5) node[pos=0.5] {PEMFC};
  \draw[black, very thick] (2.75,-1.5) rectangle (4.25,1) node[pos=0.5, align=center] {DC Link};
  \draw[black, very thick] (5.5,0.5) circle (0.5cm) node {MG};
  \draw[black, very thick] (5,-1.5) rectangle (6.5,-0.5) node[pos=0.5] {Auxiliary};
  \draw[>=triangle 45, <->] (2,0.5) -- (2.75,0.5); 
  \draw[>=triangle 45, ->] (2,-1) -- (2.75,-1); 
  \draw[>=triangle 45, <->] (4.25,0.5) -- (5,0.5); 
  \draw[>=triangle 45, ->] (4.25,-1) -- (5,-1); 
  \end{tikzpicture}
\caption{Fuel cell series hybrid architecture. Power converters and inverters are omitted for the sake of brevity. Arrows depict directions of power flow.}
\label{fig:fc_powertrain}
\end{figure}

This section presents the mathematical models used to formulate the optimization problems, namely the train's longitudinal dynamics and the powertrain's energy consumption. Figure \ref{fig:fc_powertrain} shows the fuel cell series hybrid powertrain considered. The components considered herein are the battery, fuel cell, motor, and auxiliary loads (hotel loads). The term motor is used interchangeably with motor-generator (MG).

All models are to be derived in the discrete spatial domain with a grid of $N$ sampling instances and a spatial sampling interval of $\Delta_\text{s}$ meters. The zero-order hold is to be used between sampling instances. The choice to model the system in the spatial domain instead of the typically used time domain (temporal sampling) facilitates formulating a simpler convex optimization problem \cite{RN862}.

\subsection{Longitudinal Dynamics}

Assume the train as a point mass $m$ with an equivalent inertial mass $\vartwo{m,eq}=(1+\lambda)m$, a reasonable assumption for the shorter trains investigated herein \cite{RN891}. The train's longitudinal velocity $v$ is controlled using the traction motor force $\vartwo{F,m}$ and mechanical brake force $\vartwo{F,brk}$. The external forces acting on the train $\vartwo{F,ext}$ are described by the summation of the Davis Equation and gravitational pull
\begin{equation}
\varthree{F,ext,i} = \underbrace{a+bv_i+cv_i^2}_\text{Davis Eqn.} + \underbrace{mg\sin(\theta_i)}_\text{Grav. pull}.
\end{equation}

Using the aforementioned forces, the definition of kinetic energy $E=1/2 \vartwo{m,eq} v^2$, the definition of mechanical work $E=F\Delta_\text{s}$, and the principle of energy conservation, construct
\begin{equation}\label{eq:kinetic_energy_1}
\frac{1}{2}\vartwo{m,eq}v_{i+1}^2 = \frac{1}{2}\vartwo{m,eq}v_{i}^2 + (\varthree{F,m,i} + \varthree{F,brk,i})\Delta_{\text{s},i} - \varthree{F,ext,i}\Delta_{\text{s},i}.
\end{equation}

Equation \eqref{eq:kinetic_energy_1} is nonlinear in $v$ but can be linearized by substituting the quadratic terms $v^2$ by $z$ and keeping the non-quadratic terms $v$, namely
\begin{equation}\label{eq:kinetic_energy_2}
\frac{1}{2}\vartwo{m,eq}z_{i+1} = \frac{1}{2}\vartwo{m,eq}z_{i} + (\varthree{F,m,i} + \varthree{F,brk,i})\Delta_{\text{s},i} - \varthree{F,ext,i}\Delta_{\text{s},i}.
\end{equation}

Equation \eqref{eq:kinetic_energy_2} requires the nonconvex constraint 
\begin{equation}\label{eq:v2z}
v^2=z
\end{equation} 
to hold true at the optimal solution, which is instead replaced by the relaxed convex constraint $v^2 \leq z$ \cite{RN862}. Use \eqref{eq:kinetic_energy_2} to define the linear longitudinal dynamics function
\begin{align}
\begin{split}\vartwo{l,long}(v,z,\vartwo{F,m},\vartwo{F,brk}) := z + &\frac{2\Delta_s}{\vartwo{m,eq}}(\vartwo{F,m} + \vartwo{F,brk})\\
- &\frac{2\Delta_s}{\vartwo{m,eq}}\vartwo{F,ext}.\end{split}
\end{align}

\subsection{Traction and Fictitious Forces}\label{subsec:fictitious}

The power flow in Fig. \ref{fig:fc_powertrain} is described by 
\begin{equation}\label{eq:traction_power}
\vartwo{P,m}/\vartwo{\eta,m}(\vartwo{P,m}) + \vartwo{P,aux} = \vartwo{n,fc}\vartwo{P,fc} + \vartwo{P,batt},
\end{equation}
where $\vartwo{P,m}$ is mechanical power at the wheels, $\eta_m(\vartwo{P,m})$ is motors' efficiency at converting between electric and mechanical power, $\vartwo{P,aux}$ is auxiliary power load, $\vartwo{n,fc}$ is the number of fuel cell stacks, $\vartwo{P,fc}$ is electric power output per fuel cell stack, and $\vartwo{P,batt}$ is battery electric power output.

In order to facilitate a convex formulation in the spatial domain, longitudinal forces are to be used instead of power, thus \eqref{eq:traction_power} is divided by longitudinal velocity $v$, recall ($F=P/v$), to yield
\begin{equation}\label{eq:traction_force_equality}
\vartwo{F,m}/\vartwo{\eta,m}(\vartwo{F,m},z) + \vartwo{F,aux} = \vartwo{n,fc}\vartwo{F,fc} + \vartwo{F,batt},
\end{equation}
where motor efficiency is defined as $\vartwo{\eta,m}(\vartwo{F,m},z)$ instead of $\vartwo{\eta,m}(\vartwo{P,m})$. The forces $\vartwo{F,aux}$, $\vartwo{F,fc}$, and $\vartwo{F,batt}$, are fictitious---physically meaningless---but numerically represent the share of each towards $\vartwo{F,m}/\vartwo{\eta,m}(\vartwo{F,m},z)$. Lastly, $\vartwo{F,m}/\vartwo{\eta,m}(\vartwo{F,m},z)$ can be accurately approximated by the convex second-order polynomial $\vartwo{q,m}(\vartwo{F,m},z):=p_{00}+p_{10}z+p_{01}\vartwo{F,m}+p_{11}\vartwo{F,m}v+p_{20}z^2+p_{02}\vartwo{F,m}^2$ \cite{RN862} which yields the relaxed convex inequality constraint
\begin{equation}\label{eq:traction_force}
\vartwo{q,m}(\vartwo{F,m},z) + \vartwo{F,aux} \leq \vartwo{n,fc}\vartwo{F,fc} + \vartwo{F,batt}.
\end{equation}

\subsection{Fuel Cell Energy Consumption}\label{sec:fc_model}

\begin{figure}[!t]
\centering
\includegraphics[width=8.4cm,keepaspectratio]{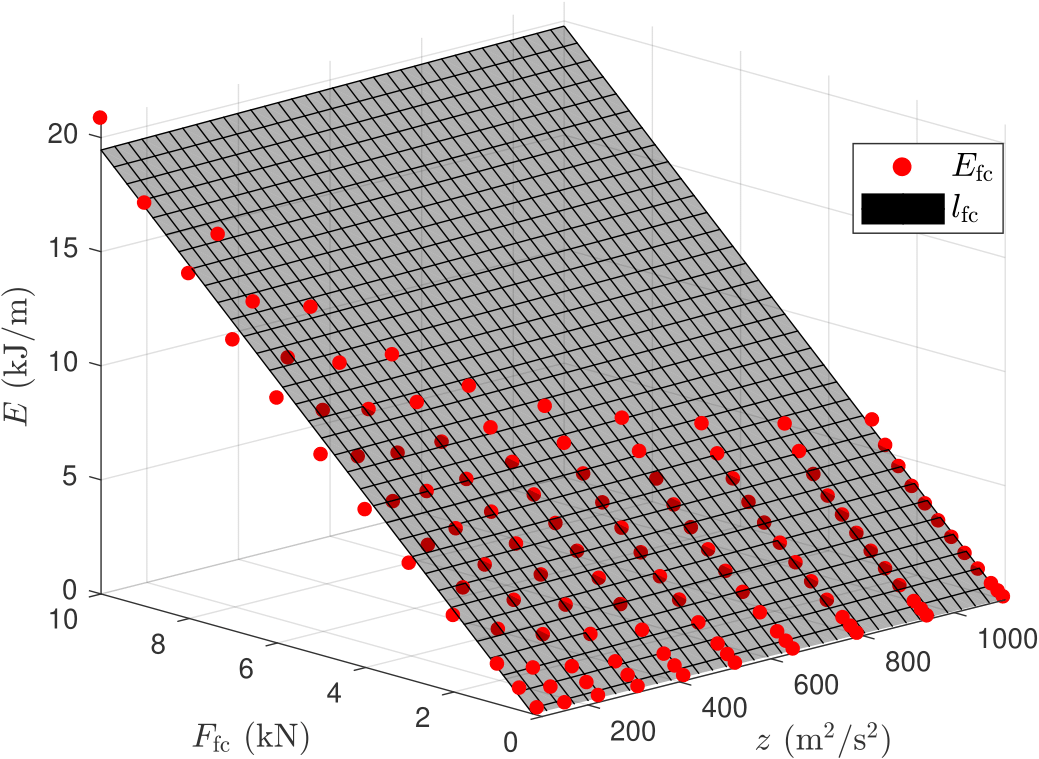}
\caption{Specific fuel cell consumption $\vartwo{E,fc}$ and linear approximation $\vartwo{l,fc}$.}
\label{fig:fc_consumption}
\end{figure}

Fuel cell efficiency is usually tabulated in a 1D look-up table against electric power output, $\vartwo{\eta,fc}(\vartwo{P,fc})$; however, it is easier to be modeled herein using $\vartwo{\eta,fc}(\vartwo{F,fc},z)$, analogous to the preference for force in section \ref{subsec:fictitious}. The hydrogen fuel energy consumed per meter traveled ($\Delta_\text{s}=1$ m) is defined by
\begin{equation}\label{eq:e_fc}
\vartwo{E,fc}(\vartwo{F,fc},z) := \frac{\vartwo{F,fc}}{\vartwo{\eta,fc}(\vartwo{F,fc},z)},
\end{equation}
which can be accurately approximated using the linear first-order polynomial $\vartwo{l,fc}(\vartwo{F,fc},z):= p_0 \vartwo{F,fc} + p_1 z$. Figure \ref{fig:fc_consumption} shows sample points of $\vartwo{E,fc}$ along the approximate hyperplane $\vartwo{l,fc}$. This approximation is fairly accurate due to the inherit linearity in the numerator of $\vartwo{E,fc}$.

\subsection{Battery State-of-Charge}\label{sec:batt_model}

Predicting the battery's state-of-charge $\zeta$ is vital in order to guarantee charge-sustaining operation. The battery considered herein is a lithium-ion battery, modeled as a fixed voltage source $\vartwo{U,oc}$ with a fixed internal resistance $R$. Experimental results from a real battery-powered train confirm the validity of such a model \cite{RN700}.

\begin{figure}[!t]
\centering
\includegraphics[width=8.4cm,keepaspectratio]{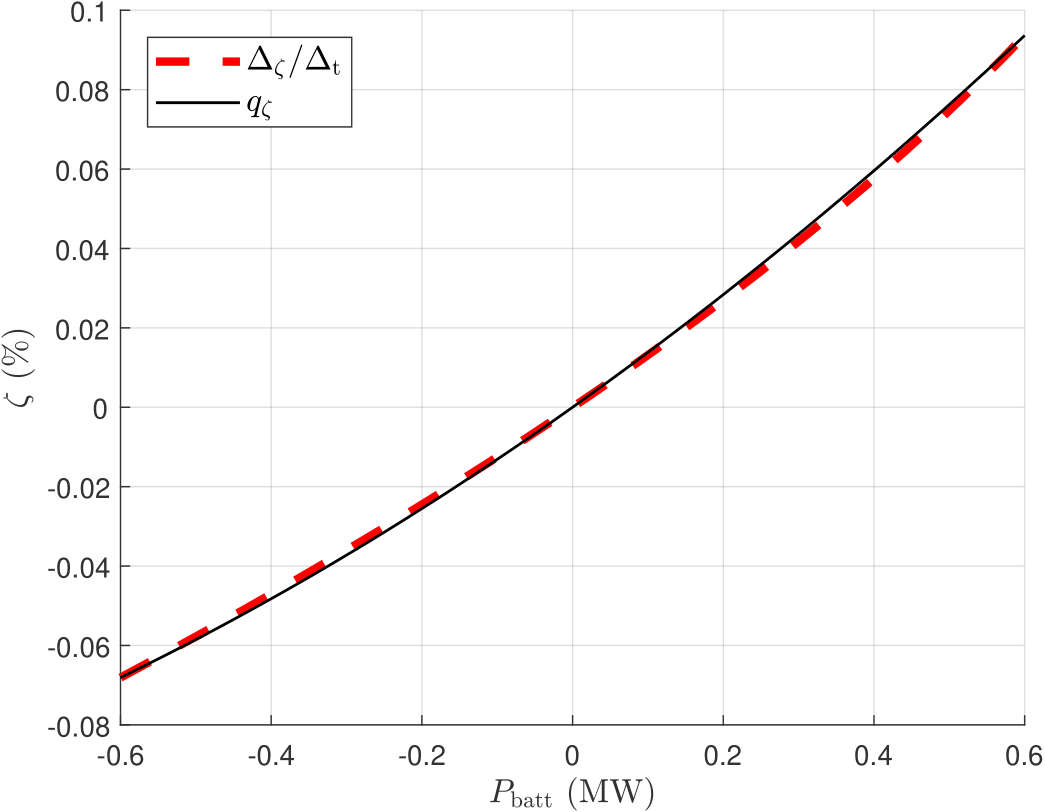}
\caption{$\Delta_\zeta/\Delta_\text{t}$ and its quadratic approximation $q_\zeta$.}
\label{fig:batt_dzeta_dt}
\end{figure}

To derive a suitable spatial domain model, start with the change in state-of-charge $\Delta_\zeta$ per discrete time step $\Delta_\text{t}$
\begin{equation}\label{eq:dzeta_dt}
\frac{\Delta_\zeta}{\Delta_\text{t}} = - \frac{\vartwo{U,oc}-\sqrt{\vartwo{U,oc}^2-4\vartwo{P,batt}R}}{2R} . \frac{1}{3600Q},
\end{equation}
where $\vartwo{P,batt}$ is electric power at battery terminal and $Q$ is battery charge capacity \cite{RN736}. A positive/(negative) value of $\Delta_\zeta$ discharges/(charges) the battery. Approximate \eqref{eq:dzeta_dt} using the convex second-order polynomial $q_\zeta(\vartwo{P,batt}):=\alpha \vartwo{P,batt}^2 + \beta \vartwo{P,batt}$. Figure \ref{fig:batt_dzeta_dt} shows how $q_\zeta$ covers $\Delta_\zeta/\Delta_\text{t}$ accurately; therefore, assuming $\Delta_\zeta/\Delta_\text{t} =q_\zeta(\vartwo{P,batt})$ leads to
\begin{equation}
\frac{\Delta_\zeta}{\Delta_\text{t}} = \alpha \vartwo{P,batt}^2 + \beta \vartwo{P,batt},
\end{equation}
which can be rewritten in terms of $\vartwo{F,batt}$ as
\begin{equation}
\frac{\Delta_\zeta}{\Delta_\text{t}} = \alpha \vartwo{F,batt}^2v^2 + \beta \vartwo{F,batt}v,
\end{equation}
followed by the substitution for $v=\Delta_\text{s}/\Delta_\text{t}$
\begin{equation}
\frac{\Delta_\zeta}{\Delta_\text{t}} = \alpha \vartwo{F,batt}^2v \frac{\Delta_\text{s}}{\Delta_\text{t}} + \beta \vartwo{F,batt} \frac{\Delta_\text{s}}{\Delta_\text{t}},
\end{equation}
cancel out $\Delta_\text{t}$ in order to obtain $\Delta_\zeta$ per spatial step $\Delta_\text{s}$ 
\begin{equation}\label{eq:battery_model_1}
\frac{\Delta_\zeta}{\Delta_\text{s}} = \alpha \vartwo{F,batt}^2v + \beta \vartwo{F,batt}.
\end{equation}

Equation \eqref{eq:battery_model_1} can be written as
\begin{equation}\label{eq:battery_model_2}
\alpha\Delta_\text{s}\vartwo{F,batt}^2 = \gamma\Omega
\end{equation}
when the bilinear constraint
\begin{equation}\label{eq:battery_model_3}
1 = v\gamma
\end{equation}
and the auxiliary variable
\begin{equation}\label{eq:battery_model_4}
\Omega=\Delta_\zeta - \beta\Delta_\text{s}\vartwo{F,batt}
\end{equation}
hold true.

The linear function $l_\zeta(\zeta,\Delta_\zeta) := \zeta - \Delta_\zeta$ can be used to predict state-of-charge after sampling interval $\Delta_\text{s}$ when used with the constraints \eqref{eq:battery_model_2}, \eqref{eq:battery_model_3}, and \eqref{eq:battery_model_4}. Both \eqref{eq:battery_model_2} and \eqref{eq:battery_model_3} need to be relaxed into the convex inequality constraints $\alpha\Delta_\text{s}\vartwo{F,batt}^2 \leq \gamma\Omega$ and $1 \leq v\gamma$, respectively, in order to obtain a convex optimization problem. Use \eqref{eq:battery_model_4} to define the auxiliary variable $\Omega$ linear function $l_\Omega(\Delta_\zeta,\vartwo{F,batt}):=\Delta_\zeta - \beta\vartwo{F,batt}\Delta_\text{s}$.

\section{Optimization Formulations}

Section \ref{sec:concurrent} presents a concurrent formulation for optimizing speed and EMS that is similar to the automotive example \cite{RN886} but with modifications relevant to the railway, most notably fuel consumption is penalized with an equality constraint on target journey time $\tau$. The proposed modifications allow the formulation to find the optimal solution under less restrictive conditions than previous relaxed formulations \cite{RN776,RN886}. The formulation is inherently nonconvex due to the nonlinear equality constraints \eqref{eq:v2z}, \eqref{eq:traction_force_equality}, \eqref{eq:battery_model_2}, and \eqref{eq:battery_model_3}, but is \textit{convexified} by relaxing them into inequality constraints.

Section \ref{sec:sequential} presents a sequential formulation that first optimizes train speed without complete knowledge of the hybrid powertrain or the EMS, followed by optimizing the EMS using the generated speed profile. This approach resembles an EMS that follows a speed profile planned by a train operator who lacks specialist knowledge of the hybrid powertrain. The sequential method is used herein as a benchmark and is similar to others from literature \cite{RN912,RN613,RN909}.

\subsection{Concurrent Method}\label{sec:concurrent}

Without loss of generality, assume all $\vartwo{n,fc}$ fuel cell stacks follow the same command $\vartwo{P,fc}=\vartwo{F,fc}v$. The cost function
\begin{equation}\label{eq:cost_function}
\sum_i \vartwo{n,fc}\vartwo{l,fc}(\varthree{F,fc,i-1},z_{i-1})\Delta_{\text{s},i-1} + \gamma_i^2 + \Omega_i^2
\end{equation}
penalizes fuel cell consumption and auxiliary variables over $i=1,\hdots,N$ sampling instances, subject to the linear equality constraints on speed squared, state-of-charge, and the auxiliary variable $\Omega$
\begin{subequations}
\begin{alignat}{1}
z_i & = \vartwo{l,long}(v_{i-1},z_{i-1},\varthree{F,m,i-1},\varthree{F,brk,i-1}),\\
\zeta_i & = l_\zeta(\zeta_{i-1},\Delta_{\zeta,i-1}),\\
\Omega_i & = l_\Omega(\Delta_{\zeta,i-1},\varthree{F,batt,i-1}),
\end{alignat}
\end{subequations}
the linear equality constraint on target journey time
\begin{equation}\label{eq:target_time}
\sum_i (\gamma_{i-1} \Delta_{\text{s},i-1}) = \tau,
\end{equation}
the terminal condition on speed and state-of-charge
\begin{subequations}
\begin{alignat}{1}
z_N & = \vartwo{z,stop},\\
\zeta_N & = \zeta_0, \label{eq:zeta_charge_sustain}
\end{alignat}
\end{subequations}
the speed constraint during station dwell time $j=1,\hdots,M$
\begin{align}
z_j & = \vartwo{z,stop}, \label{eq:dwell}
\end{align}
the lower and upper bounds
\begin{subequations}
\begin{alignat}{2}
\underline{v}_i & \leq \quad v_i && \leq \overline{v}_i,\\
\underline{z}_i & \leq \quad z_i && \leq \overline{z}_i,\\
\underline{\zeta}_i & \leq \quad \zeta_i && \leq \overline{\zeta}_i,\\
\underline{F_\text{m}}_{i-1} & \leq \varthree{F,m,i-1} && \leq \overline{F_\text{m}}_{i-1}, \label{eq:f_m_force_bounds}\\
\underline{F_\text{brk}}_{i-1} & \leq \varthree{F,brk,i-1} && \leq 0,\\ 
0 & \leq \quad \gamma_i,\\ 
0 & \leq \quad \Omega_i, 
\end{alignat}
\end{subequations}
the powertrain linear inequality constraints
\begin{subequations}
\begin{alignat}{2}
\underline{P_\text{m}}\gamma_{i-1} & \leq\ \varthree{F,m,i-1} && \leq \overline{P_\text{m}}\gamma_{i-1},\label{eq:f_m_power_bounds}\\ 
\underline{P_\text{batt}}\gamma_{i-1} & \leq \varthree{F,batt,i-1} && \leq \overline{P_\text{batt}}\gamma_{i-1},\\
\underline{P_\text{fc}}\gamma_{i-1} & \leq\ \varthree{F,fc,i-1} && \leq \overline{P_\text{fc}}\gamma_{i-1},\label{eq:p_fc}
\end{alignat}
\end{subequations}
and the relaxed nonlinear inequality constraints
\begin{subequations}
\begin{alignat}{1}
\begin{split}\vartwo{q,m}(\varthree{F,m,i-1},z_{i-1}) &+ \varthree{P,aux,i-1}\gamma_{i-1} \\ &\leq \varthree{F,fc,i-1} + \varthree{F,batt,i-1},\end{split} \label{con:traction_force} \\
v_i^2 &\leq z_i, \label{con:one} \\
1 &\leq v_i\gamma_i, \label{con:two} \\
\alpha\Delta_s\varthree{F,batt,i-1}^2 &\leq \Omega_i\gamma_{i-1}. \label{con:three}
\end{alignat}
\end{subequations}

The proposed formulation is convex and can be cast into a convex second-order cone program \cite{RN862,boyd2004convex}.

The constraint \eqref{eq:target_time} guarantees that the train completes the journey exactly $\tau$ seconds after start. The constraint \eqref{eq:zeta_charge_sustain} insures battery charge-sustained operation. The two constraints \eqref{eq:f_m_force_bounds} and \eqref{eq:f_m_power_bounds} imposed on $\vartwo{F,m}$ are the motor's force and power limits, respectively. Without loss of generality, $\underline{\vartwo{P,fc}}$ in \eqref{eq:p_fc} is strictly positive in order to prohibit fuel cell idling which is known to hamper fuel cell lifetime \cite{RN223}.

The constraints \eqref{con:one} and \eqref{con:two} imply that $z$ remains strictly positive. In order to approximate being stationary at station stops $j=1,\hdots,M$, the dwell constraint \eqref{eq:dwell} forces $z$ to a small positive value that approaches zero ($\vartwo{z,stop} \approx 0$). The external forces $\varthree{F,ext,j}$ are zeroed in order to also zero $\varthree{F,m,j}$ and $\varthree{F,brk,j}$ while approximately stationary at station stops (see \eqref{eq:kinetic_energy_2}). Since the optimized speed profile is strictly positive, the sampling intervals during station stops, $\Delta_{\text{s},j}$, are adjusted \textit{a priori} according to $\vartwo{z,stop}$ to reflect the planned dwell duration. Although the optimized speed at station stops never attains zero, in practice, it can be zeroed without affecting feasibility or optimality if $\vartwo{z,stop}$ was close enough to zero.

In order to prove the optimality of the proposed formulation, the relaxed constraints \eqref{con:traction_force}, \eqref{con:one}, \eqref{con:two}, and \eqref{con:three}, need to hold with equality at the resulting solution. The following reasoning justifies this:

\begin{enumerate}
\item \eqref{con:traction_force} holds with equality because $\vartwo{F,fc}$ and $\vartwo{F,batt}$ are pushed down to minimize fuel consumption and sustain battery charge or obtain free charge;
\item \eqref{con:two} holds with equality because \eqref{eq:cost_function} pushes down on $\gamma$ which in turn pushes up on $v$ but \eqref{eq:kinetic_energy_2} pushes down on $v$ to reduce longitudinal losses in $\vartwo{F,ext}$;
\item \eqref{con:one} holds with equality because \eqref{eq:cost_function} pushes down on $z$ while $v^2$ has been fixed by \eqref{con:two};
\item \eqref{con:three} holds with equality because the right hand side of its original form \eqref{eq:battery_model_1} pushes down in order to sustain battery charge or obtain free charge, furthermore \eqref{eq:cost_function} pushes down on both $\Omega$ and $\gamma$.
\end{enumerate}

\subsection{Sequential Method}\label{sec:sequential}

The first step to the sequential method is to obtain a reference speed profile. The formulation presented in section \ref{sec:concurrent} is modified to optimize speed while assuming ideal motors and without knowledge of batteries or fuel cells. Only the variables $v,z,\vartwo{F,m},\vartwo{F,brk},\gamma$ are optimized during this step. The variables $\vartwo{F,fc},\vartwo{F,batt},\zeta,\Delta_\zeta,\Omega$ and their respective constraints are dropped. The cost function $\sum_i \varthree{F,m,i-1}\Delta_{\text{s},i-1} + \gamma_i^2$ is used.

The second step optimizes the EMS according to the reference speed profile from the first step. The formulation from section \ref{sec:concurrent} is used with an equality constraint on the pre-generated speed profile. Only the variables $\vartwo{F,fc},\vartwo{F,batt},\zeta,\Delta_\zeta,\Omega$ are optimized. The cost function \eqref{eq:cost_function} is used.

\section{Simulation Results}

The concurrent method is compared against the sequential method by optimizing and simulating the operation of a four-car-train similar to the \textit{HydroFLEX} concept \cite{hydroflex}. In order to guarantee a fair comparison, an identical initial and terminal state-of-charge of $50\%$ was simulated. Further train parameters are shown in Table \ref{tab:train_parameters}. 

\begin{table}[!t]
\setlength{\tabcolsep}{3.5pt} 
\renewcommand{\arraystretch}{1.7} 
\caption{Simulated Train Parameters}
\label{tab:train_parameters}
\centering
\begin{tabular}{rl||rl||rl||rl}
\hline
\multicolumn{2}{l}{Vehicle}&\multicolumn{2}{l}{Motor}&\multicolumn{2}{l}{Battery}&\multicolumn{2}{l}{Fuel Cell}\\
\hline
$m$&$\SI{183}{\kilo\tonne}$&$\underline{\vartwo{P,m}}$&$\SI{-585}{\kW}$&$\underline{\vartwo{P,batt}}$&$\SI{-600}{\kW}$&$\underline{\vartwo{P,fc}}$&$\SI{6}{\kW}$\\
$\lambda$&$0.0625$&$\overline{\vartwo{P,m}}$&$\SI{585}{\kW}$&$\overline{\vartwo{P,batt}}$&$\SI{600}{\kW}$&$\overline{\vartwo{P,fc}}$&$\SI{100}{\kW}$\\
$a$&$\SI{1,743}{\kN}$&$\underline{\vartwo{F,m}}$&$\SI{-87}{\kN}$&$Q$&$\SI{220}{\kWh}$&$\vartwo{n,fc}$&$4$\\
$b$&$\SI{76.4}{\kilogram\per\second}$&$\overline{\vartwo{F,m}}$&$\SI{87}{\kN}$&$R$&$\SI{88.5}{\mohm}$& & \\
$c$&$\SI{6.2}{\kilogram\per\meter}$& & &$\underline{\zeta}$&$20$& & \\
$\underline{\vartwo{F,brk}}$&$\SI{-180}{kN}$& & &$\overline{\zeta}$&$80$& & \\
$\vartwo{P,aux}$&$\SI{100}{\kW}$& & & & & & \\
\hline
\end{tabular}
\end{table}

The rail line simulated is the $63$-km-long \textit{Tees Valley Line}, located in northern \textit{England}. The line runs between \textit{Saltburn} and \textit{Bishop Auckland} with 16 intermediate stops. The journey requires 87 minutes which was used for the journey time constraint \eqref{eq:target_time}. Both methods abide by identical dwell duration at stations but have the freedom to alter arrival and departure times of intermediate stations. Route elevation data has been extracted from \cite{elevation} using the EU-DEM dataset. The route is discretized, optimized, and simulated, at a spatial sampling interval $\Delta_\text{s}$ of 10 m, leading to around 6300 sampling instances. Despite the problem's size, the convex formulations proposed were solved on the order of tens of seconds using the barrier method \cite{gurobi}.

\begin{figure}[!t]
\centering
\includegraphics[width=8.4cm,keepaspectratio]{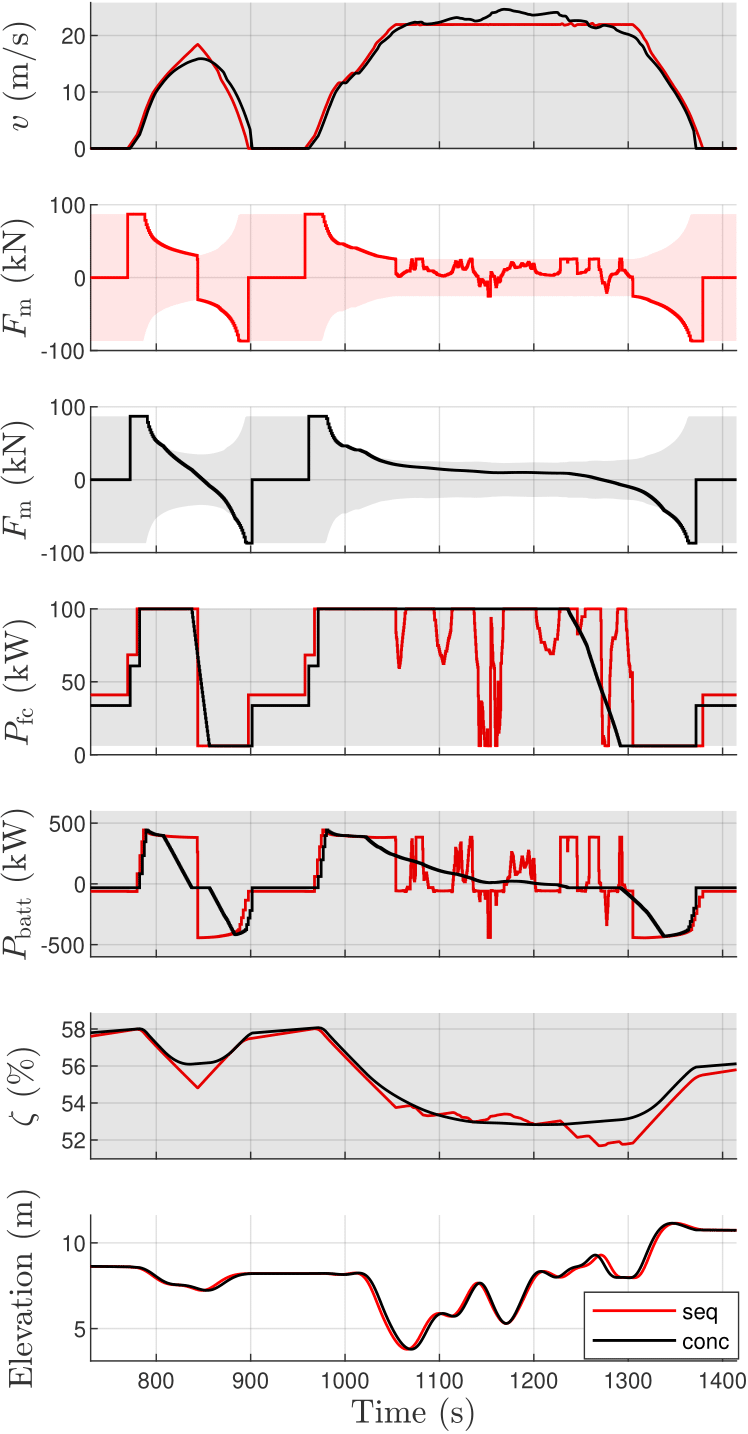}
\caption{Excerpt from simulation results. Red/(black) lines depict results of the sequential/(concurrent) method. Shaded areas depict the feasible solution set. Traction force subplots are separated because the feasible set is speed dependent. The subplots share the same time x-axis.}
\label{fig:results}
\end{figure}

Simulation results indicate that the concurrent method consumed 5\% less fuel than the sequential method. Figure \ref{fig:results} shows sample trajectories from both methods spanning three station stops. The figure showcases consistent differences in behavior that contributed towards the fuel consumption discrepancy and is further elaborated upon in the following sections.
 
\subsection{Unnecessary Regenerative Braking}\label{sec:results_1}

Figure \ref{fig:results} shows the sequential method applying regenerative braking without the intention of slowing down at $\SI{1150}{s}$ and $\SI{1275}{s}$. A closer look at the speed and elevation subplots reveals that the speed planning step of the sequential method (first step) was trying to recoup kinetic energy from local dips in elevation. Given the powertrain's double losses of regenerative braking followed by immediate positive traction, these commands are unnecessary and wasteful. The sequential method's speed planning step took such a wasteful decision because it was unaware of these losses. The concurrent method rarely exhibits such wasteful behavior, albeit none in Fig. \ref{fig:results}, instead, it largely reserves regenerative braking commands for slowing down due to its knowledge of these double losses at the time of speed planning. Furthermore, this discrepancy in behavior forces the sequential method to peak battery power and vary fuel cell power more aggressively, e.g., between $\SI{1100}{s}$ and $\SI{1300}{s}$, potentially leading to accelerated wear. Similar behavior was reported by \cite[Fig. 9]{RN613}.

\subsection{Gradual Transition to Regenerative Braking}\label{sec:results_2}

The traction force subplots in Fig. \ref{fig:results} reveal that the concurrent method gradually transitions from positive traction to regenerative braking prior to and while slowing down, whereas the sequential method abruptly switches to maximum regenerative braking, e.g., $\SI{850}{s}$. This difference is explained by the concurrent method's knowledge of powertrain efficiency characteristics while planning speed and thus chooses a more efficient speed profile that consumes less energy and recoups kinetic energy more effectively. This is confirmed by the concurrent's method higher state-of-charge graph. The sequential method's abrupt regenerative braking command is akin to the supposedly optimal solution that assumes an ideal powertrain \cite{RN707}, indeed the case for the sequential method's speed planning step.

\subsection{Speed Profile Overview}\label{sec:results_3}

\begin{figure}[!t]
\centering
\includegraphics[width=8.4cm,keepaspectratio]{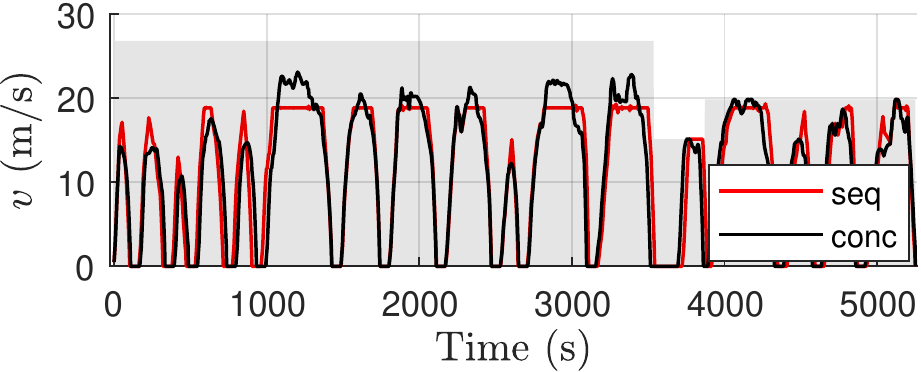}
\caption{Entire journey speed profile. Shaded area depicts speed constraints.}
\label{fig:velocity}
\end{figure}

Figure \ref{fig:velocity} shows optimized speed for both methods over the entire journey. It shows that the concurrent method tends to peak at lower velocities between closer stations, whereas it cruises at higher velocities between distant stations. The former is explained by the gradual deceleration profile preferred by the concurrent method (see section \ref{sec:results_2}), whereas the latter is explained by the need to compensate for time lost at slower speeds in order to fulfill the journey time constraint \eqref{eq:target_time}.

\section{Conclusion}
Computationally light formulations for optimizing the speed and EMS of fuel cell hybrid trains were presented. 
The benefit from concurrently optimizing all trajectories in a single optimization problem in comparison to a sequential method was demonstrated by the means of a simulation.

The concurrent method was able to find a more optimal solution than the sequential method due to its holistic knowledge of the powertrain as well as the freedom to adjust both trajectories concurrently. The sequential method optimized speed based on incomplete knowledge which initially yielded a sub-optimal speed profile. Subsequently, the sequential EMS optimization step could only find a sub-optimal EMS solution as well.

Given the crucial role played by knowledge of the powertrain, it is recommended to include further powertrain characteristics in future works, e.g., battery temperature constraints. Another avenue for investigation is to optimize the operation of each fuel cell stack independently.


\bibliographystyle{IEEEtran}
\bibliography{IEEEabrv,vppc_2021}

\end{document}